\documentclass[letterpaper,12pt]{article}
\pdfoutput=1
\usepackage{fullpage}
\usepackage{graphicx}
\usepackage{amssymb}
\usepackage{siunitx}
\usepackage[colorlinks=true,
            urlcolor=blue,
            linkcolor=blue,
            citecolor=blue]{hyperref}
\usepackage{lineno}
\usepackage[T1]{fontenc}
\begin{document} 

\begin{center}
\Large\textbf{%
Quantum atmosphere effective radii for different spin fields
from quantum gravity inspired black holes
}
\end{center}

\centerline{Douglas M. Gingrich}

\begin{center}
\textit{%
Department of Physics, University of Alberta, Edmonton, AB T6G 2E1 Canada\\
\smallskip
TRIUMF, Vancouver, BC V6T 2A3 Canada
}
\end{center}

\begin{center}
e-mail:
\href{mailto:gingrich@ualberta.ca}{gingrich@ualberta.ca}
\end{center}


\begin{abstract}
\noindent
Quantum atmosphere effective radii for the emission of
spin-0, -1/2, -1, and -2 massless fields from Schwarzschild,
Tangherlini, non-commutative geometry inspired, and polymeric black
holes are calculated.
The power observed from the black hole at spatial infinity taking
greybody factors into account is compared to an equal-power black-body
radiator of the same temperature but different effective radius.
A large range of different radii are obtained for different spin
fields and black holes.
The equal-power black-body effective radius is not, in general, a
good proxy for the location of the quantum atmosphere.
\end{abstract}

\section{Introduction}
The Hawking radiation from evaporating black holes is thought to
originate from quantum excitations near the
horizon~\cite{PhysRevD.15.365}. 
Giddings~\cite{Giddings:2015uzr} has argued that the radiation
originates from an effective radius $r_\mathrm{A}$ outside the horizon 
radius $r_\mathrm{H}$ call the quantum atmosphere: $r_\mathrm{A} -
r_\mathrm{H} \sim r_\mathrm{H}$.  
It is of interest to test the validity of Giddings' claim.

The quantum atmosphere is the location where most of the Hawking
radiation comes from.
A few different arguments have been given for the location of the
quantum atmosphere.
The thermal wavelength of typical Hawking radiation is much larger
than the horizon size.
Heuristic arguments using a gravitational version of the Schwinger
effect for particle production by tidal forces outside the horizon
have been made~~\cite{Dey:2017yez,Ong:2020hti}.
Another reasoning uses the $(1+1)$ dimensional renormalized
stress-energy tensor~\cite{Giddings:2015uzr,Dey:2017yez,Dey:2019ugf}.  
In addition, the radius can be given by an effective black-body
emission surface~\cite{Giddings:2015uzr,Hod:2016hdd}.
In this paper, we examine the later of these definitions.

Ref.~\cite{Dey:2017yez,Dey:2019ugf} have corroborated Giddings'
conclusion by obtaining $r_\mathrm{A} - r_\mathrm{H} \approx
r_\mathrm{H}$ for the Schwarzschild black hole using gravitational
Schwinger effect arguments and a more precise calculation using the
stress-energy tensor.
While the different arguments agree that the location of the quantum
atmosphere is some distance from the horizon, they do not all give a
common estimate for the numerical value. 

It is of interest to examine if Giddings' arguments are applicable to
other types of black holes.
Hod~\cite{Hod:2016hdd} showed that the quantum atmosphere radius for
a massless scalar field from a Tangherlini black hole emitting
radiation in the bulk is a decreasing function of the number of space
dimensions; Hod finds $r_\mathrm{A} - r_\mathrm{H} \ll r_\mathrm{H}$
for high number of extra dimensions.
The Reissner-Nordstr{\"o}m black hole has also been considered in
Ref.~\cite{Ong:2020hti}.
These metrics give contradicting conclusions to
Ref.~\cite{Giddings:2015uzr,Dey:2017yez,Dey:2019ugf}.

In this paper, we calculate exact greybody factors numerically
for all spin fields.
The results are used to calculate the double-differential frequency
spectrum which is then integrated over all frequencies to obtain the
power.
By equating the power to that of a black body, we determine an
effective emission surface of the quantum atmosphere.
The potentials seen by different spin fields are different so we could 
expect the quantum atmosphere to depend on the emitted field's spin.
We find that the apparent radius should not be used, in
general, as a proxy for the location of the quantum atmosphere.
For example, $r_\mathrm{A}$ can not be used as a definition for the
location of the quantum atmosphere for gravitons for most black hole
metrics we consider. 

%
\section{Effective radius calculation}
The effective potential barrier around a black hole is commonly
encoded in a set of transmission coefficients, greybody factors, that
depend on the properties of the black hole, the properties of the
emitted radiation, frequency and modes of the emitted radiation.
A physical observable that can be formed from the transmission
coefficients is the absorption cross section which is a sum of the
transmission coefficients over all radiation modes divided by the
frequency squared.
Weighting the absorption cross section by a temperature-dependent
statistical factor corresponding to the spin-statistics  of the emitted
radiation gives the radiation flux or power per unit frequency.
By integrating over all energies, the total radiated power or
luminosity is obtained.
In the absence of absorption -- step-function transmission coefficients
-- the Stefan-Boltzmann law is obtained.
For black holes, one can convert the temperature dependence into a
dependence on the horizon radius, and in principle a dependence on the
black hole parameters.
The power thus allows a determination of the effects of the
transmission coefficients integrated over all frequencies.
By comparing the power generated by a black hole with the
equivalent power from a black body seen at spatial infinity, one
obtains and effective area for the black hole, or in the case of
spherically symmetric  black holes, an effective radius.
The method of equal-power infers the size of the radiating body.

The calculated power emitted from a black hole seen by an observe at
spatial infinity is compared to the equivalent power $P_\mathrm{B}$
from an idealize black-body radiator in flat space using the
generalized Stefan-Boltzmann relation (see for example
Ref.~\cite{Cardoso})  

\begin{equation}
P_\mathrm{B}  = \sigma A_{n+2}({r_\mathrm{A}}) T_\mathrm{B}^{n+4}\, ,
\label{eq:BB}
\end{equation}

\noindent
where $T_\mathrm{B}$ is the black- body temperature,
$A_{n+2}(r_\mathrm{A})$ is the surface area of a $(n+4)$-dimensional
emitting body, and $\sigma$ is the appropriate Stefan-Boltzmann
constant for bosons or fermions in $n$ extra dimensions.
We use units of $G = c = \hbar = k_\mathrm{B} = 1$.
Although Eq.~(\ref{eq:BB}) is written in the general form to allow
comparison with higher-dimensional black holes, it reduces to the more
familiar form of the Stefan-Boltzmann law when $n=0$.

For a black hole, once the greybody factors $\Gamma_{s,\ell}(\omega)$
for massless spin field $s$ emitted with spheroidal harmonic mode
$\ell$ with frequency $\omega$ have been calculated, the absorption
cross section in four spacetime dimensions  is obtained:

\begin{equation}
\sigma_s(\omega) = \frac{\pi}{\omega^2} \sum_{\ell\ge s}
(2\ell+1) \Gamma_{s,\ell}(\omega)\, .
\end{equation}

\noindent
The $(2\ell+1)$ factor is the degeneracy of the axial quantum number
or angular momentum $m$ modes.

The total power in four spacetime dimensions is then given by

\begin{equation}
P = \frac{1}{2\pi^2} \int_0^\infty
\frac{\omega^3\sigma_s(\omega)}{\exp(\omega/T) -
(-1)^{2s}} \mathrm{d}\omega\, ,
\label{eq:BH}
\end{equation}

\noindent
where $T$ is the Hawking temperature as measured at spatial infinity.

We define the effective radius $r_\mathrm{A}$ of the black hole
quantum atmospheres by equating the Hawking radiation power from the
black hole Eq.~(\ref{eq:BH}) with the corresponding Stefan-Boltzmann 
radiation power of a flat space perfect black-body emitter
Eq.~(\ref{eq:BB}): 

\begin{equation}
P(r_\mathrm{H},T) = P_\mathrm{B}(r_A,T_\mathrm{B})\, .
\end{equation}

\noindent
This equation determines the effective radius assuming equal
temperature: $T = T_\mathrm{B}$. 
One could likewise determine the effective temperature of the filtered
radiation by assuming equal radii~\cite{Bekenstein:1993bg}.

Using $A_{n+2} \propto R^{n+2}$, we obtain the effective radius using 

\begin{equation}
\frac{r_\mathrm{A}}{r_\mathrm{H}}
= \left[ \frac{P(r_\mathrm{H},T)}{P_\mathrm{B}(r_\mathrm{H},T)}
\right]^\frac{1}{n+2}\, , 
\end{equation}

\noindent
where $P$ depends on the emitted field's spin and $P_\mathrm{B}$ is
different for bosons and fermions.
The dimensionless radii $r_\mathrm{A}/r_\mathrm{H}$ characterizes the
black hole quantum atmospheres.

As in Ref.~\cite{Hod:2016hdd}, it is beneficial to characterize the
effective quantum atmosphere using 

\begin{equation}
\bar{r}_\mathrm{A} = \frac{ r_\mathrm{A} - r_\mathrm{H} }{ r_\mathrm{H} }\, .
\end{equation}

\noindent
Values of $\bar{r}_\mathrm{A} \gtrsim 1$ validate Giddings' argument and
negative values imply the quantum atmosphere is behind the horizon.

%
\section{Black holes thermodynamics}
In this section, we write down the black-body power for the different
metrics consider.
The formula contain only a single polarization for each spin field.
We make no claim about the validity of the two quantum inspired black
hole metrics considered here.
They are partly chosen for their different black-body features and the
ease of greybody calculation.

\subsection{Schwarzschild-Tangherlini black holes}

For the Schwarzschild-Tangherlini~\cite{tangherlini1963schwarzschild}
black hole radiating into the bulk, the higher dimensional $(n+4)$
black-body power is~\cite{Cardoso,Hod:2016hdd}

\begin{equation}
P_\mathrm{B}  = \sigma A_{n+2}(R) T^{n+4}\, ,
\end{equation}

\noindent
where the higher-dimensional Stefan-Boltzmann constant is

\begin{equation}
\sigma  = \frac{(n+3)\Gamma((n+3)/2) \zeta(n+4)}{2\pi^{(n+3)/2+1}}\, ,
\end{equation}

\noindent
and $\Gamma$ is the gamma function and $\zeta$ is the Riemann zeta
function.
The higher-dimensional surface area of the emitting body of radius $R$
is 

\begin{equation}
A_{n+2}(R)  = \frac{2\pi^{(n+3)/2}}{\Gamma((n+3)/2)} R^{n+2}\, .
\end{equation}

\noindent
We will also need the black hole temperature

\begin{equation}
T  = \frac{n+1}{4\pi r_\mathrm{H}}\, ,
\end{equation}

\noindent
where

\begin{equation}
r_\mathrm{H} = \frac{1}{\sqrt{\pi} M_*} \left( \frac{M}{M_*}
\right)^{1/(n+1)} \left[ \frac{8\Gamma((n+3)/2)}{n+2}
\right]^{1/(n+1)}\, .
\end{equation}

\noindent
The above equations reduce to the familiar Stefan-Boltzmann law
and Schwarzschild black hole when $n=0$, and $M_* = \sqrt{\hbar c/G}$
is the Planck mass. 

\subsection{Non-commutative geometry inspired black holes}

Non-commutative geometry inspired black holes are interesting in that
the form of the black-body area of the Schwarzschild-Tangherlini
remains unchanged but the temperature dependence is
different~\cite{Nicolini:2005vd,Rizzo:2006zb}. 
The temperature is given by

\begin{equation}
T  = \frac{n+1}{4\pi r_\mathrm{H}} \left[ 1 - \frac{2}{n+1}
  \left( \frac{r_\mathrm{H}}{2\sqrt{\theta}} \right)^{n+3}
  \frac{\mathrm{e}^{-r_\mathrm{H}/(4\theta)}}{\gamma\left(\frac{n+3}{2},
    \frac{r_\mathrm{H}^2}{4\theta}\right)} \right]\, ,
\end{equation}

\noindent
where $\gamma$ is the upper incomplete gamma function.
The horizon radius is obtained by solving

\begin{equation}
\frac{M}{M_*} = \frac{n+2}{8\gamma\left(\frac{n+3}{2},
  \frac{r^2}{4\theta} \right)} (\sqrt{\pi}M_* r_\mathrm{H})^{n+1}\, .
\end{equation}

\noindent
The minimum length parameter $\sqrt{\theta}$ is take to be a free
parameter and could be well above the Planck length.
As $\theta \to 0$, the radius and temperature approach the Tangherlini
values.
The metric give one, two, or no horizon.
For a single horizon the temperature vanishes and a black hole remnant
is expected to form. 
The temperature has a maximum but vanishes at the remnant radius.
The non-commutative black hole is similar to the Tangherlini black
hole for large masses.

To model the effects of an effective ultra-violet cut-off in the
frequency $\omega$ of the emitted quanta an additional
factor~\cite{Nicolini:2011nz} of $\exp(-\theta\omega^2/2)$
should multiply Eq.~(\ref{eq:BH}).
Although we have included this factor, it has a small effect.

\subsection{Polymeric black holes}

In loop quantum gravity, semi-classical corrections due to the
effects of quantum gravity have been derived to give a so-called
polymer Schwarzschild black
hole~\cite{Modesto:2009ve,Modesto:2008im}. 
The model has two free parameters $\epsilon$ and $a_0$.
The parameter $a_0 = 8\pi A_\mathrm{min}$ is related to the minimum
area of loop quantum gravity and is expected to be of the Planck scale. 

A positive deformation parameter $\epsilon$ represents the
typical scale of the geometry fluctuations in the Hamiltonian
constraints of the theory as they get renormalized from the Planck
scale to the astrophysical scales.
It's thought that $\epsilon \ll 1$, and values of $\epsilon \lesssim 0.8$ 
will have little effect on what follows.
For large $\epsilon$, deviations from the Schwarzschild metric are
apparent for astronomical size black holes.

The horizon area is not the usual form but is given by

\begin{equation}
A = 4\pi (2m)^2 \left[ 1
  + \left( \frac{\sqrt{a_0}}{2m} \right)^4 \right]\, .
\end{equation}

\noindent
The temperature is given by

\begin{equation}
T = \frac{1}{4\pi(2m)} (1-P(\epsilon)^2) \left[ 1 + \left(
    \frac{\sqrt{a_0}}{2m}
    \right)^4 \right]^{-1}\, ,
\end{equation}

\noindent
where the polymerization function is

\begin{equation}
P(\epsilon) = \frac{\sqrt{1+\epsilon^2}-1}{\sqrt{1+\epsilon^2}+1}\, .
\end{equation}

\noindent
The total integrated power given by the Stefan-Boltzmann law is

\begin{equation}
P = \frac{\sigma}{256\pi^3} m^{-2} (1-P(\epsilon)^2)^4
\left[ 1 + \left( \frac{\sqrt{a_0}}{2m}
  \right)^4\right]^{-3}\, ,
\end{equation}

\noindent
where $\sigma = \pi^2/120$ for bosons and $\sigma = 7\pi^2/960$ for
fermions.
In the above equations $m$ is a parameter that is related to the ADM
mass $M$ by $M = m(1+P)^2$.

%
\section{Results}
We calculate the quantum atmosphere effective radius for spin-0, -1/2,
-1, and -2 massless fields from two quantum inspired black holes. 
Our calculations are numerical and follow the procedures used in
Ref~\cite{Cox:2023lal} which are based on the general potentials in
Ref~\cite{Arbey:2021jif} and the path-ordered matrix exponentials in
Ref~\cite{Gray:2015xig}. 
The procedure enables previously rather difficult calculations.

\subsection{Schwarzschild black hole}

We consider the Schwarzschild black holes as a warm-up.
Table~\ref{tab:S} shows dimensionless effective radii for all spin
fields from a Schwarzschild black hole.
Our numerical calculations reproduce the results of
Page~\cite{Page:1976df} for spin-1, -1/2, -2, and
Elster~\cite{ELSTER1983205} for scalars.
In terms of the quantum atmosphere, the case of spin-1 was first discussed
in Ref.~\cite{Giddings:2015uzr} and the spin-0 in
Ref.~\cite{Hod:2016hdd}. 
The case of spin-2 shows a breakdown of Gidding's principle
(Ref.~\cite{Giddings:2015uzr} restricted the discussion to $s \le 1$). 

\begin{table}[htb]
\centering
\caption{\label{tab:S}%
Dimensionless radii $\bar{r}_\mathrm{A}$ for massless fields of spin $s$ from
a Schwarzschild black hole.  
}
\smallskip
\begin{tabular}{ccccc}\hline
$s$ & 0 & 1/2 & 1 & 2\\\hline
$\bar{r}_\mathrm{A}$ & 1.68 & 1.13 & 0.27 & $-0.57$\\
\hline
\end{tabular}
\end{table}

For the Schwarzschild black hole, the black-body power is well known to
have a $P \sim M^{-2}$ dependence.
We find that including graybody factors, this mass dependence is
maintained, i.e.\ $\Gamma$  does not introduce any additional $M$
dependence.

\subsection{Tangherlini black hole}

To help validate our procedure, we reproduce a previous result in
Ref.~\cite{Hod:2016hdd}. 
Table~\ref{tab:ST_bulk} shows dimensionless effective radii for
scalars from a Tangherlini black hole radiating in the bulk. 
We have taken $M = M_* = 1$.
To obtain these results, we have calculated the emission on the brane
and used the bulk-to-brane emission ratios obtained in
Ref.~\cite{Harris:2003eg}. 
Our results agree with Ref.~\cite{Hod:2016hdd} to within the numerical
accuracy of the calculations.

\begin{table}[htb]
\centering
\caption{%
Dimensionless radii $\bar{r}_\mathrm{A}$ for a massless scalar field from
a $(n+4)$-dimensional Tangherlini black hole radiating in the bulk. 
}
\label{tab:ST_bulk}
\smallskip
\begin{tabular}{cccccccc}\hline
$n$ & 1 & 2 & 3 & 4 & 5 & 6 & 7\\\hline
$\bar{r}_\mathrm{A}$ & 0.99 & 0.71 & 0.59 & 0.50 & 0.44 & 0.39 & 0.33\\
\hline
\end{tabular}
\end{table}

We are now equipped to calculate something new.
Table~\ref{tab:ST_brane} shows dimensionless effective radii for all spin
fields from a Tangherlini black hole radiating on the brane.
Looking at the large values of $\bar{r}_\mathrm{A}$ for brane emission,
we reach a different conclusion from 
bulk emission, and support Giddings' argument much better.

\begin{table}[htb]
\centering
\caption{%
Dimensionless radii $\bar{r}_\mathrm{A}$ for massless fields of spin $s$ from
a $(n+4)$-dimensional Tangherlini black hole radiating on the brane. 
}
\label{tab:ST_brane}
\smallskip
\begin{tabular}{cccccccc}\hline
& \multicolumn{7}{c}{$n$}\\
$s$  & 1 & 2 & 3 & 4 & 5 & 6 & 7\\\hline
0   & 3.44 & 4.98 & 5.83 & 5.86 & 5.21 & 4.16 & 2.78\\
1/2 & 3.44 & 5.10 & 5.90 & 5.84 & 5.13 & 4.04 & 2.71\\
1   & 2.68 & 4.70 & 5.82 & 5.99 & 5.39 & 4.32 & 2.95\\
2   & 0.96 & 2.67 & 3.89 & 4.37 & 4.14 & 3.43 & 2.34\\
\hline
\end{tabular}
\end{table}

\subsection{Non-commutative geometry inspired black hole}

The non-commutative geometry inspired black hole we consider has a
minimum horizon radius at a finite mass (a black hole remnant), and a
temperature that has a maximum before the temperature vanishes.
Thus the power does not follow the $M^{-2}$ dependence near the end of
the black hole's lifetime and $\bar{r}_\mathrm{A}$ depends on the
black hole mass. 
For high $M\sqrt{\theta}$, we reproduce the Schwarzschild results.
For the black-body case, below about $M\sqrt{\theta} < 6$, the power
dependence deviates from a pure $M^{-2}$ dependence and vanishes as
$M\to 1.9/\sqrt{\theta}$.
The black hole power falls faster than the black-body power with $M$
except for the spin-0 field. 

Table~\ref{tab:NC} shows dimensionless effective radii for all spin
fields from a non-commutative geometry inspired black hole in higher
dimensions radiating on the brane at the maximum temperature; we have
taken $\sqrt{\theta} = 1$. 

\begin{table}[htb]
\centering
\caption{%
Dimensionless radii $\bar{r}_\mathrm{A}$ for massless fields of spin $s$ from
a $(n+4)$-dimensional non-commutative geometry inspired black hole
radiating on the brane with the maximum temperature and $\sqrt{\theta}
= 1$.
}
\label{tab:NC}
\smallskip
\begin{tabular}{cSccccccc}\hline
& \multicolumn{8}{c}{$n$}\\
$s$  & \multicolumn{1}{c}{0} & 1 & 2 & 3 & 4 & 5 & 6 & 7\\\hline
0 & 1.70    & $-0.98$ & 5.85 & 7.35 & 7.93 & 7.62 & 6.60 & 5.21\\
1/2 & 1.03    & $-0.98$ & 6.03 & 7.51 & 7.99 & 7.57 & 6.47 & 5.04\\
1 & 0.12    & $-0.99$ & 5.33 & 7.22 & 8.04 & 7.83 & 6.83 & 5.42\\
2 & -0.68 & $-0.99$ & 2.75 & 4.50 & 5.55 & 5.77 & 5.25 & 4.28\\
\hline
\end{tabular}
\end{table}
\subsection{Polymeric black hole}

The polymeric black hole also has a maximum temperature but the
temperature vanishes at zero mass.
Combined with the non-trivial area dependence, the power does not
follow a $M^{-2}$ dependence and $\bar{r}_\mathrm{A}$ depends on the
mass of the black hole.
We have taken $\epsilon = 0.01$ and $a_0 = 1$.
For this value of $\epsilon$, $P = 2.5\times 10^{-5}$ and gives a
negligible contribution to the power, and causes $m \approx M$.
For high $2M/\sqrt{a_0}$, we reproduce the Schwarzschild results.
For the black-body case, below about $2M/\sqrt{a_0} < 2$, the power
dependence deviates from a pure $M^{-2}$ dependence and vanishes as
$M\to 0$.
The black hole power falls faster than the black-body power with $M$
except for the spin-0 field. 

Table~\ref{tab:LQG} shows dimensionless effective radii for all spin
fields from a polymeric black hole at the maximum temperature.

\begin{table}[htb]
\centering
\caption{%
Dimensionless radii $\bar{r}_\mathrm{A}$ for massless fields of spin $s$ from a
polymeric black hole with the maximum temperature, and $\epsilon =
0.01$ and $a_0 = 1$. 
}
\label{tab:LQG}
\smallskip
\begin{tabular}{ccccc}\hline
$s$ & 0 & 1/2 & 1 & 2\\\hline
$\bar{r}_\mathrm{A}$ & 1.57 & 0.62 & $-0.22$ & $-0.90$\\
\hline
\end{tabular}
\end{table}

%
\section{Discussion}
We have calculated the quantum atmosphere for all massless spin fields
for the first time.
Two quantum gravity inspired metrics posing different black-body
power formula have been compared with exact numerical calculations
of the total power from the black hole including greybody factors.

Giddings' argument of $\bar{r}_\mathrm{A} \sim 1$ clearly depends on
the spin of the emitted radiation, decreasing by a factor of about six
when going from scalars to vectors, and in general does not apply to 
gravitons.

Hod's~\cite{Hod:2016hdd} result $\bar{r}_\mathrm{A} < 1$ for
higher-dimensional black holes is reproduced, but if the radiation is  
confined to our brane, the conclusion is very different.
Values of $\bar{r}_\mathrm{A} \sim 5$ for most spins and
extra dimensions are obtained.
The higher-dimensional form of the black-body formula plays a
significant role beyond just the greybody factors.

We have examined two quantum gravity inspired black holes in the
regime were quantum effects are important and the radiation will have
its maximum intensity.
The quantum atmosphere for scalar fields in four space-time dimensions
appears similar regardless of the quantum inspired metric and is
similar to Schwarzschild black holes.  

The power in the spin-0 field always has a quantum atmosphere radius
of about 1.7 times the horizon radius in four space-time dimensions.
We can see that, in general, the effective radius of an
equivalent black-body radiator is not a good proxy for the quantum
atmosphere.
On the other hand, the effective radius $\bar{r}_\mathrm{A}$ could be
considered an intuitive measure of greybody effects on the total power
received by an observer.

The greybody factors themselves are of little interest until they are
used to calculate physical observables.
It is common to calculate the absorption cross section and compare the
high-frequency limit against the geometric cross section and the
low-frequency limit against the surface area.
These limits allow an easily quantifiable measure of the greybody
effects of different metrics.
Perhaps a more measurement observable, someday, will be the total
particle fluxes and energy spectra measured by a distant observer.
First measurements of these quantities are likely to be integrated
over the detecting instrument's acceptance and resolution to obtain
single numbers for the number of particles per unit time and energy
per unit time (or power), before full spectra are measured.
Expressing these measurements in terms of an effective black-body
radius could prove to be a useful mnemonic for elucidating quantum gravity
effects. 

\section{Acknowledgments}

We acknowledge the support of the Natural Sciences and Engineering
Research Council of Canada (NSERC). 
Nous remercions le Conseil de recherches en sciences naturelles et en
g{\'e}nie du Canada (CRSNG) de son soutien. 
\bibliographystyle{JHEP}
\bibliography{gingrich}

\providecommand{\href}[2]{#2}\begingroup\raggedright\begin{thebibliography}{10}

\bibitem{PhysRevD.15.365}
W.G.~Unruh, \emph{Origin of the particles in black-hole evaporation},
  \href{https://doi.org/10.1103/PhysRevD.15.365}{\emph{Phys. Rev. D} {\bfseries
  15} (1977) 365}.

\bibitem{Giddings:2015uzr}
S.B.~Giddings, \emph{{Hawking radiation, the Stefan\textendash{}Boltzmann law,
  and unitarization}},
  \href{https://doi.org/10.1016/j.physletb.2015.12.076}{\emph{Phys. Lett. B}
  {\bfseries 754} (2016) 39}
  [\href{https://arxiv.org/abs/1511.08221}{{\ttfamily 1511.08221}}].

\bibitem{Dey:2017yez}
R.~Dey, S.~Liberati and D.~Pranzetti, \emph{{The black hole quantum
  atmosphere}},
  \href{https://doi.org/10.1016/j.physletb.2017.09.076}{\emph{Phys. Lett. B}
  {\bfseries 774} (2017) 308}
  [\href{https://arxiv.org/abs/1701.06161}{{\ttfamily 1701.06161}}].

\bibitem{Ong:2020hti}
Y.C.~Ong and M.R.R.~Good, \emph{{Quantum atmosphere of Reissner-Nordstr\"om
  black holes}},
  \href{https://doi.org/10.1103/PhysRevResearch.2.033322}{\emph{Phys. Rev.
  Res.} {\bfseries 2} (2020) 033322}
  [\href{https://arxiv.org/abs/2003.10429}{{\ttfamily 2003.10429}}].

\bibitem{Dey:2019ugf}
R.~Dey, S.~Liberati, Z.~Mirzaiyan and D.~Pranzetti, \emph{{Black hole quantum
  atmosphere for freely falling observers}},
  \href{https://doi.org/10.1016/j.physletb.2019.134828}{\emph{Phys. Lett. B}
  {\bfseries 797} (2019) 134828}
  [\href{https://arxiv.org/abs/1906.02958}{{\ttfamily 1906.02958}}].

\bibitem{Hod:2016hdd}
S.~Hod, \emph{{Hawking radiation and the Stefan\textendash{}Boltzmann law: The
  effective radius of the black-hole quantum atmosphere}},
  \href{https://doi.org/10.1016/j.physletb.2016.03.071}{\emph{Phys. Lett. B}
  {\bfseries 757} (2016) 121}
  [\href{https://arxiv.org/abs/1607.02510}{{\ttfamily 1607.02510}}].

\bibitem{Cardoso}
T.~Cardoso and A.~Castro, \emph{The blackbody radiation in a {D}-dimensional
  universes},
  \href{https://doi.org/10.1590/S0102-47442005000400007}{\emph{Revista
  Brasileira de Ensino de Física} {\bfseries 27} (2005) }.

\bibitem{Bekenstein:1993bg}
J.D.~Bekenstein, \emph{{How fast does information leak out from a black
  hole?}}, \href{https://doi.org/10.1103/PhysRevLett.70.3680}{\emph{Phys. Rev.
  Lett.} {\bfseries 70} (1993) 3680}
  [\href{https://arxiv.org/abs/hep-th/9301058}{{\ttfamily hep-th/9301058}}].

\bibitem{tangherlini1963schwarzschild}
F.R.~Tangherlini, \emph{Schwarzschild field in n dimensions and the
  dimensionality of space problem},
  \href{https://doi.org/https://doi.org/10.1007/BF02784569}{\emph{Il Nuovo
  Cimento (1955-1965)} {\bfseries 27} (1963) 636}.

\bibitem{Nicolini:2005vd}
P.~Nicolini, A.~Smailagic and E.~Spallucci, \emph{{Noncommutative geometry
  inspired Schwarzschild black hole}},
  \href{https://doi.org/10.1016/j.physletb.2005.11.004}{\emph{Phys. Lett. B}
  {\bfseries 632} (2006) 547}
  [\href{https://arxiv.org/abs/gr-qc/0510112}{{\ttfamily gr-qc/0510112}}].

\bibitem{Rizzo:2006zb}
T.G.~Rizzo, \emph{{Noncommutative inspired black holes in extra dimensions}},
  \href{https://doi.org/10.1088/1126-6708/2006/09/021}{\emph{JHEP} {\bfseries
  09} (2006) 021} [\href{https://arxiv.org/abs/hep-ph/0606051}{{\ttfamily
  hep-ph/0606051}}].

\bibitem{Nicolini:2011nz}
P.~Nicolini and E.~Winstanley, \emph{{Hawking emission from quantum gravity
  black holes}}, \href{https://doi.org/10.1007/JHEP11(2011)075}{\emph{JHEP}
  {\bfseries 11} (2011) 075} [\href{https://arxiv.org/abs/1108.4419}{{\ttfamily
  1108.4419}}].

\bibitem{Modesto:2009ve}
L.~Modesto and I.~Premont-Schwarz, \emph{{Self-dual black holes in LQG: Theory
  and phenomenology}},
  \href{https://doi.org/10.1103/PhysRevD.80.064041}{\emph{Phys. Rev. D}
  {\bfseries 80} (2009) 064041}
  [\href{https://arxiv.org/abs/0905.3170}{{\ttfamily 0905.3170}}].

\bibitem{Modesto:2008im}
L.~Modesto, \emph{{Semiclassical loop quantum black hole}},
  \href{https://doi.org/10.1007/s10773-010-0346-x}{\emph{Int. J. Theor. Phys.}
  {\bfseries 49} (2010) 1649}
  [\href{https://arxiv.org/abs/0811.2196}{{\ttfamily 0811.2196}}].

\bibitem{Cox:2023lal}
Z.~Cox and D.M.~Gingrich, \emph{{Greybody factors for higher-dimensional
  non-commutative geometry inspired black holes}},
  \href{https://arxiv.org/abs/2303.08309}{{\ttfamily 2303.08309}}.

\bibitem{Arbey:2021jif}
A.~Arbey, J.~Auffinger, M.~Geiller, E.R.~Livine and F.~Sartini, \emph{{Hawking
  radiation by spherically-symmetric static black holes for all spins:
  Teukolsky equations and potentials}},
  \href{https://doi.org/10.1103/PhysRevD.103.104010}{\emph{Phys. Rev. D}
  {\bfseries 103} (2021) 104010}
  [\href{https://arxiv.org/abs/2101.02951}{{\ttfamily 2101.02951}}].

\bibitem{Gray:2015xig}
F.~Gray and M.~Visser, \emph{{Greybody factors for Schwarzschild black holes:
  Path-ordered exponentials and product integrals}},
  \href{https://doi.org/10.3390/universe4090093}{\emph{Universe} {\bfseries 4}
  (2018) 93} [\href{https://arxiv.org/abs/1512.05018}{{\ttfamily 1512.05018}}].

\bibitem{Page:1976df}
D.N.~Page, \emph{{Particle emission rates from a black hole: Massless particles
  from an uncharged, nonrotating hole}},
  \href{https://doi.org/10.1103/PhysRevD.13.198}{\emph{Phys. Rev. D} {\bfseries
  13} (1976) 198}.

\bibitem{ELSTER1983205}
T.~Elster, \emph{Vacuum polarization near a black hole creating particles},
  \href{https://doi.org/https://doi.org/10.1016/0375-9601(83)90449-8}{\emph{Physics
  Letters A} {\bfseries 94} (1983) 205}.

\bibitem{Harris:2003eg}
C.M.~Harris and P.~Kanti, \emph{{Hawking radiation from a (4+n)-dimensional
  black hole: Exact results for the Schwarzschild phase}},
  \href{https://doi.org/10.1088/1126-6708/2003/10/014}{\emph{JHEP} {\bfseries
  10} (2003) 014} [\href{https://arxiv.org/abs/hep-ph/0309054}{{\ttfamily
  hep-ph/0309054}}].

\end{thebibliography}\endgroup
\end{document}